# Simultaneous monitoring of the two coupled motors of a single $F_oF_1$-ATP synthase by three-color FRET using duty cycle-optimized triple-ALEX

N. Zarrabi[a], S. Ernst[a], M. G. Düser[a], A. Golovina-Leiker[a], W. Becker[b], R. Erdmann[c], S. D. Dunn[d], M. Börsch*[a]

[a]3rd Institute of Physics, University of Stuttgart, Pfaffenwaldring 57, 70550 Stuttgart, Germany;
[b]Becker&Hickl GmbH, Nahmitzer Damm 30, 12277 Berlin, Germany
[c]PicoQuant GmbH, Rudower Chaussee 29, 12489 Berlin, Germany
[d]Department of Biochemistry, University of Western Ontario, London, Ontario, Canada N6A 5C1

## ABSTRACT

$F_oF_1$-ATP synthase is the enzyme that provides the 'chemical energy currency' adenosine triphosphate, ATP, for living cells. The formation of ATP is accomplished by a stepwise internal rotation of subunits within the enzyme. Briefly, proton translocation through the membrane-bound $F_o$ part of ATP synthase drives a 10-step rotary motion of the ring of $c$ subunits with respect to the non-rotating subunits $a$ and $b$. This rotation is transmitted to the γ and ε subunits of the $F_1$ sector resulting in 120° steps. In order to unravel this symmetry mismatch we monitor subunit rotation by a single-molecule fluorescence resonance energy transfer (FRET) approach using three fluorophores specifically attached to the enzyme: one attached to the $F_1$ motor, another one to the $F_o$ motor, and the third one to a non-rotating subunit. To reduce photophysical artifacts due to spectral fluctuations of the single fluorophores, a duty cycle-optimized alternating three-laser scheme (DCO-ALEX) has been developed. Simultaneous observation of the stepsizes for both motors allows the detection of reversible elastic deformations between the rotor parts of $F_o$ and $F_1$.

## 1. INTRODUCTION

We study the conformational dynamics of the enzyme $F_oF_1$-ATP synthase which catalyzes the formation of adenosine triphosphate, ATP, from ADP and phosphate. $F_oF_1$-ATP synthases are large membrane proteins consisting of at least 8 different subunits. They are located in the plasma membrane of bacteria, in the inner mitochondrial membrane and in the thylakoid membrane of chloroplasts. According to the chemiosmotic theory[1] the driving force to synthesize ATP is the difference of the electrochemical potential of protons across a lipid membrane, that is, the enzyme from the bacterium *Escherichia coli* utilizes a pH difference plus an electric potential generated by ion concentration differences to make ATP. Proton translocation through the membrane-bound $F_o$ part of ATP synthase drives a rotary motion of the ring of $c$ subunits with respect to the non-rotating subunits $a$ and $b$ (Fig. 1). This rotation is transmitted to the γ and ε subunits of the $F_1$ sector which accommodates three catalytic sites for binding ATP or ADP plus phosphate. These sites are located at the interface of the β and α subunits. Rotation of the γ and ε subunits induces conformational changes in the binding sites which favor binding of the substrates, catalytic activity and release of the products. The rotary motion of γ synchronizes the highly cooperative and sequential catalytic processes in each of the three binding pockets [2].

..................................................................................................................................................................

*m.boersch@physik.uni-stuttgart.de; phone (49) 711 6856 4632; fax (49) 711 6856 5281; www.m-boersch.org

Direct experimental evidence for this "binding change mechanism" suggested by P. Boyer has been obtained from crystallographic structural information[3], biochemical methods[4] and by spectroscopic means[5]. The conclusive demonstration of subunit rotation during catalysis was accomplished by single-molecule observation. The bacterial $F_oF_1$-ATP synthases can switch to ATP hydrolysis, and thereby the γ and ε subunits in $F_1$ (as well as the coupled ring of c subunits in $F_o$) are forced to rotate in a direction opposite to that observed during ATP synthesis. In 1997 the laboratories of M. Yoshida and K. Kinosita used this reversibility of the chemical reaction, that is, ATP hydrolysis mode, to show rotating μm-long actin filaments which were attached to the γ subunit of surface-immobilized $F_1$ fragments[6]. Subsequently the videomicroscopic approach was refined by using smaller marker beads to reduce the viscous drag. Following the unraveling of substeps of γ subunit rotation[7] the application of the $F_1$ part as a biological rotary nanomotor in bionanotechnology was proposed and evaluated. The rotation mechanism of ε was confirmed *in vivo* by fusions of globular proteins of various sizes to either its N or C terminus[8, 9]; large fusion constructs prevented subunit ε to pass the peripheral b subunit dimer and blocked rotation: Thus, $F_oF_1$-ATP synthase-dependent *E. coli* growth was stopped.

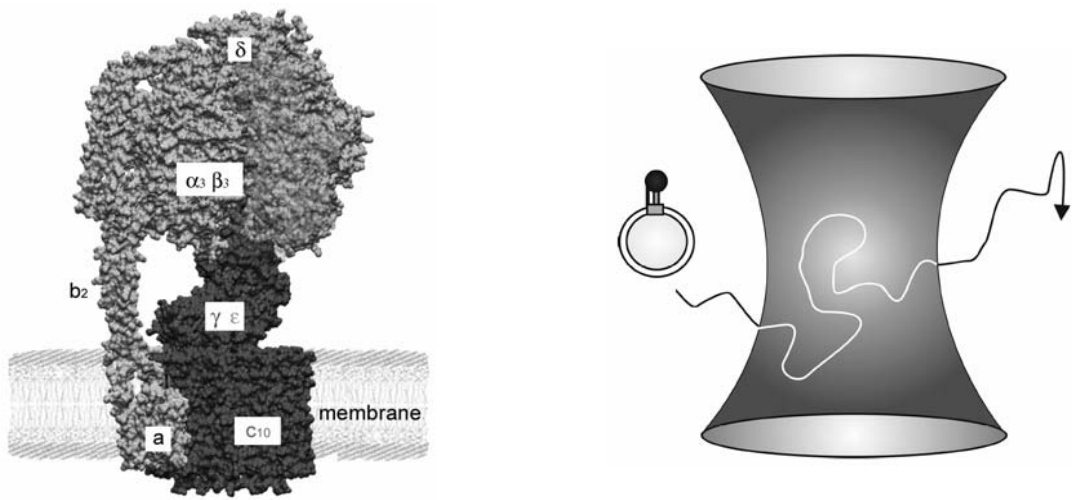

Fig. 1. Left, model of $F_oF_1$-ATP synthase embedded in the lipid membrane of a vesicle. The lipid membrane separates two buffer solutions with different pH for proton-driven ATP synthesis. An additional electric potential across the membrane can be generated using different $K^+$ concentrations. The $F_1$ motor consists of the non-rotating subunits $\alpha_3\beta_3\delta$ (grey) and the rotating γ and ε subunits (black). The $F_o$ motor comprise the static subunits *a* and $b_2$ (grey) and the rotating ring of ten *c* subunits (black). Right, confocal FRET approach to monitor rotation in single liposome-bound $F_oF_1$-ATP synthases, which are freely diffusing in buffer solution through the laser focus.

We have developed an alternative fluorescence approach to monitor conformational changes and rotary subunit movements in single membrane-embedded $F_oF_1$-ATP synthase from E. coli. Freely diffusing lipid vesicles with a diameter of 100 - 150 nm containing a single enzyme molecule are used at essentially the same conditions as in the biochemical assays to measure ATP synthesis and hydrolysis rates (Fig 1b). Through this procedure, any influence of surfaces on the catalytic activities, for instance hindered diffusion of the nucleotides ATP, ADP or phosphate, or a slowdown on rotation caused by viscous drag on a large attached reporter probe, are avoided. We apply single-molecule Förster-type fluoresence resonance energy transfer (FRET) using two fluorophores to monitor the rotary motions [10-18]. The well-known distance dependence of FRET is widely used to map the spacing between two intramolecular or intermolecular positions in the range between 2 and 8 nm. We can resolve distance changes between the two fluorophores, one covalently attached to one of the rotating subunits and the second one bound to a non-rotating subunit, in real time with millisecond time resolution and sub-nanometer precision. Subunit rotation will result in a sequence of distance changes between these two fluorophores, and, the FRET efficiency is expected to change in a congruent, stepwise manner.

We started to monitor γ subunit rotation (γ labeled at a cysteine genetically introduced at residue position 106) by FRET using a non-rotating labeling position at a specific lysine on the β subunit of the $F_1$ sector. This lysine can be labeled selectively with sulfonyl fluoride derivatives of rhodamines at pH 9.0 and 4°C [19]. The expected distances between the two fluorophores were large (in the range of 10 nm, see Fig. 2b). However we observed three distinct FRET efficiencies in the presence of AMPPNP. Because AMPPNP is a non-hydrolyzable ATP derivative, it blocks rotation of the γ subunit and traps γ in one angular orientation. During ATP hydrolysis at 1 mM ATP we found sequential changes of three FRET efficiencies. The sequence of FRET levels →1→2→3→1→ supported the interpretation that we were observing distance changes due to rotation of the γ subunit within single active $F_oF_1$-ATP synthases. Upon ATP synthesis conditions, the sequence of FRET levels was reversed [20] indicating the opposite direction of rotation during ATP synthesis and ATP hydrolysis.

The signal-to-noise ratio for the single-molecule FRET experiments with $F_oF_1$-ATP synthase could be improved by replacing the FRET donor fluorophore with tetramethylrhodamine, TMR. The distances changes and the related FRET efficiency changes were enlarged by attaching the second fluorophore to the b subunits of the $F_o$ part (at cysteines introduced to residue 64, see Fig. 2c). Comparing continuous-wave excitation with ps pulsed laser excitation we could reveal that the measured FRET efficiency changes based on fluorescence intensities were due to FRET and not due photophysical artifacts of fluorophores in the local protein environment of residue 106 at γ [21].

Similarly we revealed 120° stepping of the ε subunit during ATP hydrolysis at 1 mM ATP (see Fig. 2e). For ATP synthesis, again 120° stepping was observed but in the opposite direction [13, 18, 22]. Small differences in the dwell times for the three stopping positions of ε were attributed to the asymmetry induced by the peripheral stalk of the two b subunits. We assume that the protein interactions might affect the conformational dynamics of the opening and closing motions in at least one α/β subunit pair. The differences in dwell times corresponded to the asymmetric FRET efficiency histograms, or to distributions of stopping positions of ε, in the absences of added nucleotides or in the presence of AMPPNP, respectively.

This asymmetry of dwell times was further used to triangulate the C terminus of the proton-translocating subunit a in $F_o$. by FRET [14, 15, 23] (Fig. 2 g). Therefore, an autofluorescent protein, EGFP, was fused to the C terminus via a short amino acid linker. A single EGFP was bright enough for single-molecule detection and for subsequent intramolecular FRET measurements. However, limited photostability and laser power-dependent spectral fluctuations and shifts restricted the use of EGFP as a FRET donor label.

One way to avoid the photobleaching-limited observation times in single-molecule fluorescence detection is to permanently exchange the fluorophores. It has been noticed for many FRET fluorophore pairs the FRET acceptor is more often photobleached before the FRET donor is destroyed. Accordingly we used an fluorescently labeled ATP derivative as an exchangeable FRET acceptor (Fig. 2d). The binding pocket can partly accommodate the large fluorophore and the ATP gets hydrolyzed at slower rates [24]. The single-molecule FRET experiment to monitor rotation of γ during ATP hydrolysis revealed more than three FRET levels (or γ subunit orientations, respectively) with respect to one nucleotide binding site, indicating the existence of additional stopping positions for γ [25]. This has been observed also for isolated $F_1$ fragments at low ATP concentrations using a small bead of 40 nm and high speed videomicroscopy [7].

For the proton-driven c ring rotation in the $F_o$ part of ATP synthase, the step size of rotation might be expected to correspond to the passage of each c subunit past the proton-translocating a subunit. However, rotation of the c ring during ATP hydrolysis using detergent-solubilized $F_oF_1$-ATP synthase was found to occur in large 120° steps [26], while smooth rotation was observed using either actin filaments[27] or a single fluorophore attached to a $Na^+$-driven ATP synthase[28]. To apply our single-molecule FRET approach to c ring rotation we fused the FRET donor fluorophore EGFP to the a subunit and introduced one cysteine in each of the 10 c subunits for substoichiometric labeling with the FRET acceptor (Fig. 2i). For ATP synthesis conditions in the presence ADP, phosphate and a pH gradient plus an electrical potential across the membrane, we clearly identified rotational steps of the c ring that were smaller than 120°.

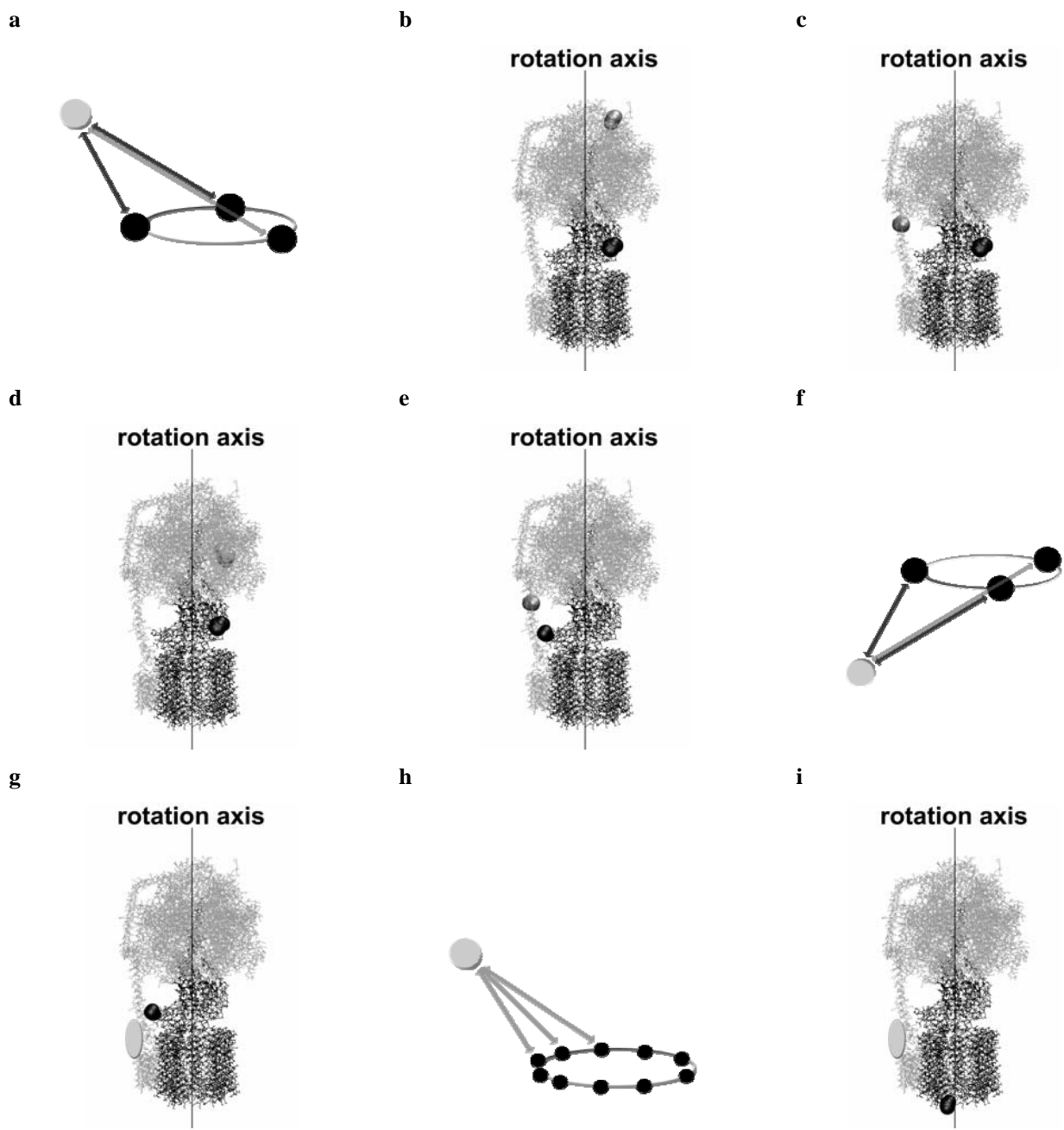

Fig. 2. Expected FRET distance changes for rotating (black dots) versus static (grey dots) labeling positions and the corresponding positions in $F_oF_1$-ATP synthase models. **a**, three distinct distances according to 120° steps. **b**, FRET between labels at lysine 4 (shown as a grey dot) on β and γ106 (black dot). **c**, FRET labels at residues b64 (grey) and γ106 (black). **d**, FRET labels at a bound ATP derivative (grey) and γ106 (black). **e**, FRET labels at residues b64 (grey) and ε56 (black,. **f**, three distinct distances according to 120° steps for FRET donor at subunit *a*. **g**, FRET labels at the C terminus of subunit *a* (grey) and ε56 (black). **h**, 10 distinct distances according to 36° steps. i, FRET labels at subunit *a* (grey) and on one of 10 *c* subunits (black).

Here we show that the *c* subunit likely rotates in smaller steps (ultimately 36°) in lipid membrane-embedded $F_oF_1$ during ATP hydrolysis, in contrast to the larger steps previously reported for detergent-solubilized enzymes. Accordingly the mismatch between the 120° stepping of the γ and ε subunit and this smaller step size of the *c* ring will result in relative distance changes between the ε and the *c* subunits. To monitor this elastic deformation within the rotor part of $F_oF_1$ we developed a FRET scheme using three intramolecular marker positions simultaneously (Fig. 3). As a static reference point within $F_oF_1$ we chose the *a* subunit and fused genetically an autofluorescent protein, EGFP, to its C terminus. For the internal conformational dynamics we labeled the ε subunit with Alexa532 at ε56 and one of the *c* subunits with Cy5. With this triple-FRET scheme, the FRET efficiency fluctuations between the ε subunit and the *c* subunits can be related to an overall motion of the rotor subunits, and photophysical artifacts can be minimized.

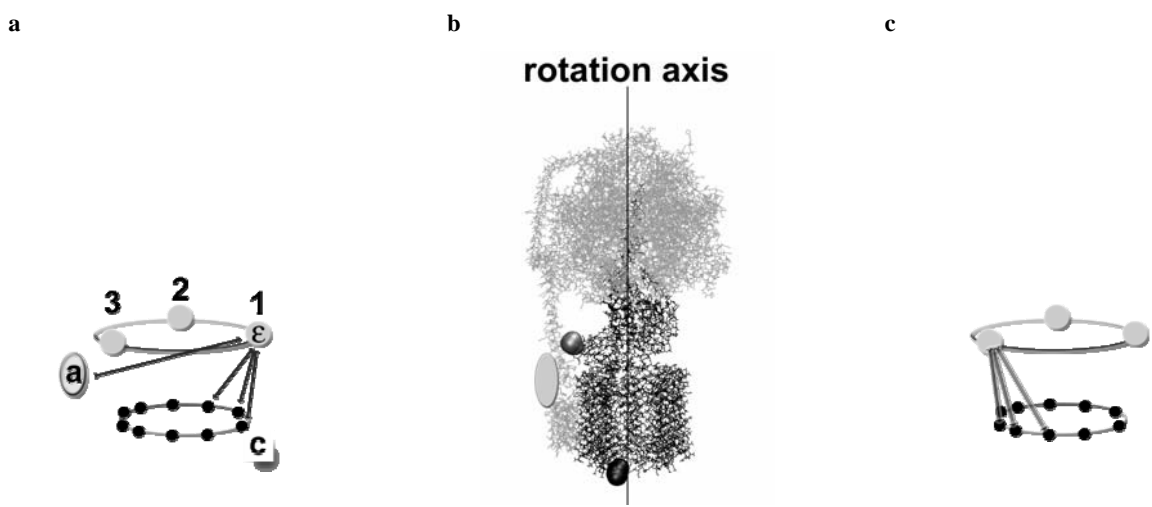

Fig. 3. Exploring the internal conformational dynamics of the rotor subunits ε and *c* of $F_oF_1$-ATP synthase by triple-FRET. **a**, possible FRET distance changes for rotating *c* subunits (black dots) versus rotating ε56 (grey round dots) with respect to the static reference position at subunit *a* (grey ellipse). **b**, labeling positions for the three fluorophores used in the triple-FRET experiment in the $F_oF_1$ model. **c**, step-by-step FRET distance changes for rotating *c* subunits (black dots) versus rotating ε56 (grey round dots) as measured by time resolved FRET using duty cycle optimized alternating laser excitation (DCO-ALEX).

## 2. EXPERIMENTAL PROCEDURES

### 2.1. Fluorescently labeled $F_oF_1$-ATP synthases for triple-FRET in liposomes

The general preparation procedures for $F_oF_1$-ATP synthase from *Escherichia coli* and the exchange of the $F_1$ part has been published. Briefly, the $F_oF_1$-ATP synthases used for single-molecule triple-FRET measurements and the FRET measurements of the internal rotor dynamics contained a genetic fusion of the 'enhanced Green Fluorescent Protein', EGFP, to the C terminus of subunit *a* of $F_o$, that has been constructed by Y. Bi (University of Western Ontario, Canada). The FRET acceptor Cy5 (GE Healthcare) was attached to a genetically introduced reactive cysteine (cys-2) at the rotating *c* subunits. Both of these fluorophores are bound to the membrane-integral $F_o$ sector. The third fluorophore was Alexa532 bound to an introduced cysteine (cys-56) in the rotating ε subunit of the $F_1$ part.

Because simultaneous labeling of the cysteines in the two different subunits cannot be done selectively in one step, we labeled the *c* ring with Cy5 first using detergent solubilized enzymes. Cy5 labeling efficiency was determined by UV-VIS absorption spectroscopy using the EGFP absorbance as an internal concentration reference. Enzymes were reconstituted in an excess of preformed liposomes (radii of about 100 nm) to assure the ratio of a single enzyme per liposome. Afterwards we exchanged the unlabeled $F_1$ part with Alexa532-labeled $F_1$. ATP hydrolysis rates were measured for each step as a control. The triple-labeled $F_oF_1$-ATP synthases in liposomes were stored as 2 or 5 μl aliquots at -80° C.

**2.2. Confocal microscope setup for triple-FRET**

The triple-FRET measurements were accomplished on a custom-designed confocal microscope based on an inverted Olympus IX 71 which had been modified compared to the previously reported excitation and detection schemes [29-43]. Fiber coupled picosecond pulsed laser sources were available at 488 nm (PicoTa 490, up to 80 MHz repetition rate, Picoquant) and 635 nm (LDH-P-635B, Picoquant) which both could be triggered externally. Continuous-wave laser excitation at 532 nm could be achieved by a frequency-doubled Nd:YAG at 532 nm (Compass 315M, Coherent) which was switched in nanoseconds using a fast acousto-optic modulator (AOM, model 3350-192, Crystal technologies). For some of the FRET experiments we temporarily replaced this laser by a new picosecond pulsed laser diode head at 530 nm (LDH-P-FA-530, up to 80 MHz repetition rate, Picoquant) which could by triggered and synchronized by an external pulse generator.

For the triple-FRET experiments, laser beams at 488 nm, 532 nm and 635 nm were compressed for the solution measurements to achieve enlarged focal spots. They were overlaid manually using two dichroic beam splitters (see Fig 4). Entering the microscope stage from the epi-fluorescence port, the beams were re-directed by a dichroic filter (triple-band beam splitter 488/532/633, AHF) and focussed into the solution by a water-immersion objective (UPlanSApo 60xW, 1.2 N.A., Olympus). Fluorescence was detected by three avalanche photodiodes simultaneously (AQR-14, Perkin Elmer) in three spectral ranges using two consecutive dichroic beam splitters (DCXR 532 and DCXR 630, AHF). Briefly, fluorescence passed a 150 μm pinhole, and EGFP was detected between 504 nm and 522 nm (HQ 513/17; AHF), Alexa532 in the second channel between 545 nm and 625 nm (HQ585/80; AHF), and Cy5 in the FRET acceptor channel between 663 nm and 737 nm (HQ700/75, AHF).

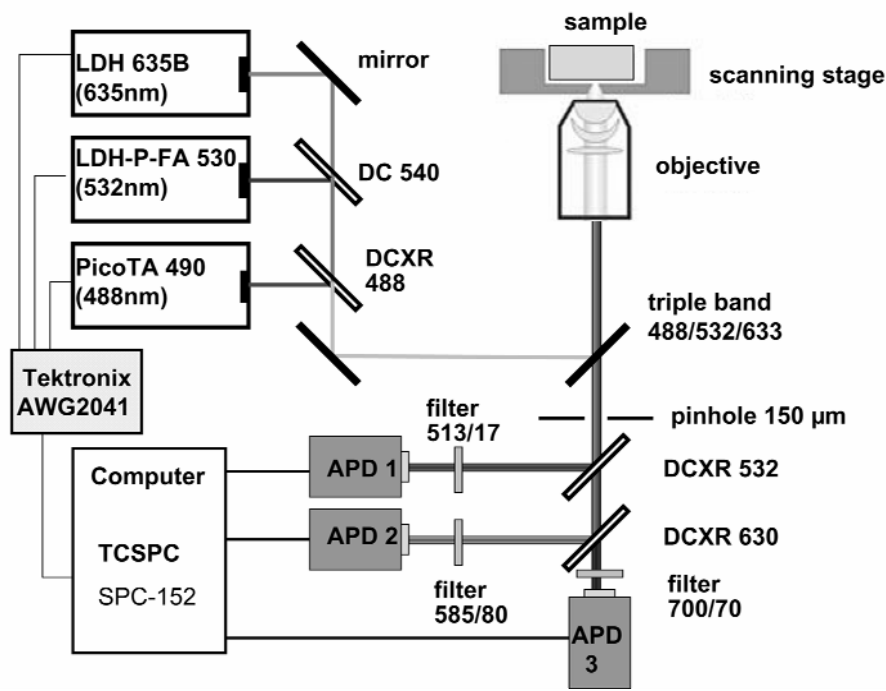

Fig. 4: Setup for the confocal time resolved FRET measurements of the internal rotor dynamics in freely diffusing single $F_oF_1$-ATP synthases. For the triple-FRET experiments an AOM-switched Nd:YAG at 532 nm was used instead of the ps pulsed LDH-P-FA 530.

Single photons were registered by a set of TCSPC cards (either two synchronized SPC-152 for 2 APDs or one SPC-150 plus router electronics for 3 APDs, Becker & Hickl) and simultaneously by a multi channel counter card for imaging (NI-PCI 6602, National Instruments). Sample scanning was accomplished by an x-y piezo scanner plus piezo objective positioner (Physik Instrumente), and the digital controller was addressed by custom-made software written in LabView. Images were analyzed with the software 'ODA-analyzer' written by N. Zarrabi.

Time trajectories of the fluorescence intensities of FRET labeled $F_oF_1$-ATP synthases were analyzed using the software package 'Burst-analyzer' written by N. Zarrabi. We used the microtime tag of each photon registered with picosecond resolution, that is, related to the synchronization trigger of the TCSPC electronics, to obtain the fluorescence lifetime for the FRET donor within a single photon burst or FRET efficiency level. Following the separation of the microtime histograms of the APDs with respect to the laser pulses we reconstructed the time trajectories for each fluorophore depending on the exciting laser pulse. Using this software gating approach we could eliminate the Raman signal from the buffer solution in the second detection channel upon pulsed excitation at 488 nm. Since the Raman signal was intense, it could otherwise have been misinterpreted as an apparent FRET efficienicy (see below). The macrotime with 50 ns resolution for each photon was used for fluorescence correlation spectroscopy and for binning to 1ms time intervals to analyze single-molecule FRET by the relative fluorescence intensities.

**2.3. Duty cycle optimized alternating laser excitation (DCO-ALEX) schemes with variable repetition rates**

One key component for the triple-FRET experiments is the Arbitrary Waveform Generator (AWG2041) used to trigger the three lasers and the photon counting electronics. The device displayed the 8-bit digital signals corresponding to the analog waveform with a time resolution of 1 ns. The digital ECL signal was transformed to TTL signals by a custom-made transformer in order to control the switch for the AOM, to trigger the pulsed laser and to synchronize the TCSCP cards. In this way we could optimize the duty cycle for each laser independently for exciting the three labels at the $F_oF_1$-ATP synthase. For instance, the narrow band interference filter used to block the 532 nm laser resulted in a weakened donor fluorescence signal for EGFP. To obtain a useful photon count rate this required a sequence of two or more consecutive ps pulses from the PicoTa 490 laser.

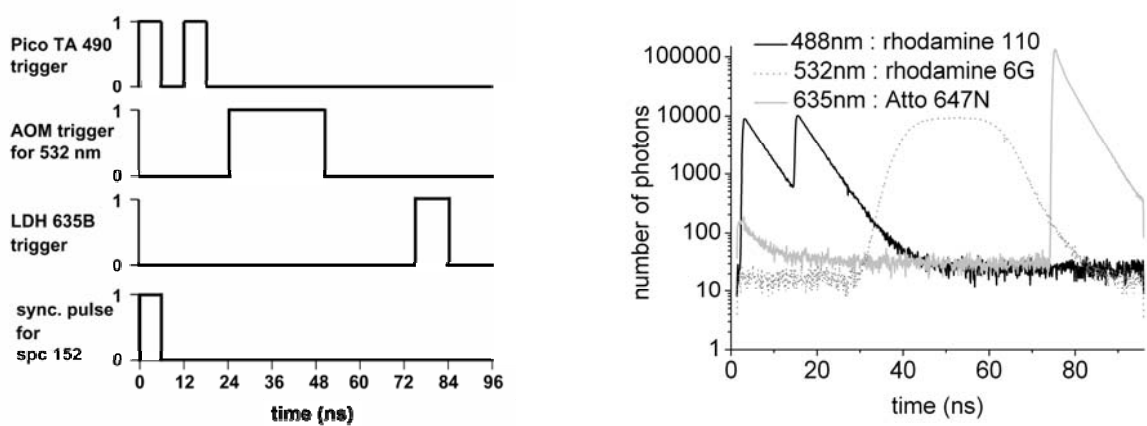

Fig. 5: Example of TTL pulse sequences from the Tektronix AWG2041 for multiple pulsed laser excitation. Left, trigger pulse sequence for the AOM, PicoTA 490 and the laser diode LDH-P-635B, and the TCSPC cards synchronization. Right, microtime histogram of the photons as detected from the three APDs in one period of 96 ns. All fluorophores were present simultaneously in solution. Only one laser was exciting the sample (the other two were blocked) while the histogram was accumulated.

For each of the three FRET experiments reported here a distinct sequence of laser pulses was applied.

1. As reported previously for monitoring *c* ring rotation with respect to subunit *a*, we optimized the photon count rates of the two fluorophores, EGFP and Alexa-568, by generating a series of six consecutive pulses for the donor-exciting laser and one pulse for the acceptor-probing laser. The 12 ns pulse to the AOM turned the acceptor-probing laser on, and the following 12 ns pause switched the 561 nm laser off. The 10 ns rising and falling times of the AOM led to the smooth shape of the first peak in the microtime histogram of the TCSPC cards. Subsequently, six consecutive pulses to the PicoTA laser with an interval of 12 ns created the six fluorescence lifetime peaks in the microtime histogram. With this pulse sequence the maximum excitation rate of the FRET donor of 83.3 MHz dropped down to only 75% (i.e. 62.5 MHz) instead of 33% (27.7 MHz) in a standard alternating approach[16].

2. For monitoring the internal relative movements between the ε and c subunits we used the new picosecond pulsed laser diode head at 530 nm LDH-P-FA-530 to excite FRET donor Alexa532. With a fluorescence lifetime of about 4 ns, the subsequent laser pulse had to be shifted by 18 ns to be able to measure the fluorescence decay for Alexa532 completely. It turned out to be sufficient to probe the FRET acceptor fluorophor Cy5 with the pulsed 635 nm laser in a simple non-symmetric alternating excitation scheme because of the short decay time of Cy5 (see Fig. 7). After having established the relationship of relative fluorescence intensity ratios of the two fluorophores and the corresponding shortening of the fluorescence lifetimes of Alexa532 due to distance-dependent FRET we continued to use the AOM-switched Nd:YAG at 532 nm for the intensity-based analysis of FRET efficiency changes during rotation of the ε subunit with respect to subunit c.

## 3. RESULTS

### 3.1. FRET measurements of single $F_oF_1$-ATP synthases labeled at subunits a and c during ATP hydrolysis

In our previous single-molecule FRET experiments (Fig. 2a – g) we have observed a 120° stepped rotation of the rotor subunits γ and ε during ATP hydrolysis at high ATP concentration as well as during ATP synthesis. In the coupled $F_oF_1$-ATP synthase, ATP hydrolysis is associated with proton transport across the lipid membrane. Current mechanistic models propose the existence of two half-channels in the proton-translocating subunit a of the $F_o$ part. A proton enters one half channel and moves towards the adjacent proton binding site on one c subunit at the interface of both subunits. As the proton binds to the empty anionic site the charge is neutralized and the ring of c subunits moves by electrostatic forces one step ahead[44]. This corresponds to a stepsize of 36° if the c ring consists of 10 subunits.

To monitor c ring rotation in a coupled $F_oF_1$-ATP synthase, that is, for the case that the enzyme is reconstituted into a lipid vesicle and is capable of ATP synthesis, we fused the autofluorescent protein EGFP to the C terminus of the a subunit. As the second fluorophore for FRET we attached Alexa568 to one c subunit (see Fig. 2i). The maleimide function of the dye reacted with the cysteine at amino acid position 2, which was introduced genetically (to be published elsewhere). Using substoichiometric amounts of the reactive dye we could achieve a very low labeling ratio with respect to the EGFP that was incorporated in each $F_oF_1$.

We reconstituted the double-labeled enzymes into liposomes with a diameter of about 150 nm and checked ATP hydrolysis and ATP synthesis rates by biochemical assays. Afterwards, the solution was diluted to an $F_oF_1$-ATP synthase concentration of about 100 pM for confocal single-molecule measurements. We used two alternating lasers at 488 nm and 561 nm to excite the FRET donor and to probe the existence of the FRET acceptor, respectively. The laser pulse sequence was optimized for a high duty cycle to excite the FRET donor. Briefly, a series of six ps pulses at 488 nm was followed by one AOM-switched pulse at 561 nm[16].

Fig. 6 a shows the photon burst of a single $F_oF_1$-ATP synthase in the presence of 1 mM ATP. In the lower panel the fluorescence intensity trajectories of FRET donor EGFP on a and of FRET acceptor Alexa568 on c were anticorrelated , indicating relative distance changes between the fluorophores, i.e. rotation of the c ring driven by ATP hydrolysis. The middle trace displays the corresponding FRET acceptor intensity trajectory upon direct excitation, confirming the existence of the Alexa568 on c. FRET efficiencies in each time interval were approximated by the proximity factor value, P,

$$P = \frac{I_A}{I_A + I_D} \qquad (1)$$

where $I_A$ and $I_D$ are the background-corrected fluorescence intensities in the FRET acceptor and FRET donor channels, respectively. In the upper panel, P, a measure of the actual FRET efficiency, indicated that the observed distance changes corresponded to a full rotation cycle of the c ring within about 60 ms.

Within this $F_oF_1$-ATP synthase burst, the proximity factor P=0.75 at an observation time of t=40 ms corresponds to a distance of about 4 nm using the previously measured Förster radius $R_0$=4.9 nm and the correction factor γ for the detection efficiencies. This spacing of the fluorophores is just the thickness of the lipid bilayer, *i.e.* it corresponds to the shortest distance that will be seen during *c*-ring rotation.

Then, the distance increased within 15 ms due to rotation towards a distal orientation, as indicated by the decline in the proximity factor of P to 0.15. Afterwards the FRET distance decreased again and reached the starting orientation with P=0.75 at t=80ms. For this photon burst, the stepwise change of distances apparently did not relate to a step size of 120° as reported for the thermophilic ATP synthase solubilized in detergent[26]. Instead, a smaller step size seemed to be associated with ATP hydrolysis. As anticipated from the mechanism of the proton-pumping process during ATP hydrolysis, *i.e.*, each *c* subunit passing the proton channel *a* one after another, a step size of 36° is to be expected. To prove that the shown trace represents a typical step size, we accumulated all measured FRET distance changes in the FRET transition density plot[45] in Fig. 6b. A significant number of FRET transitions showed small step sizes and corresponded to the theoretical ellipse (grey curve) for 36° stepping. However, we also found larger step sizes which could be interpreted as 120° stepping. It is not yet clear if these larger step sizes were due to missing short dwells caused by our limited time resolution and the manual assignment of FRET levels.

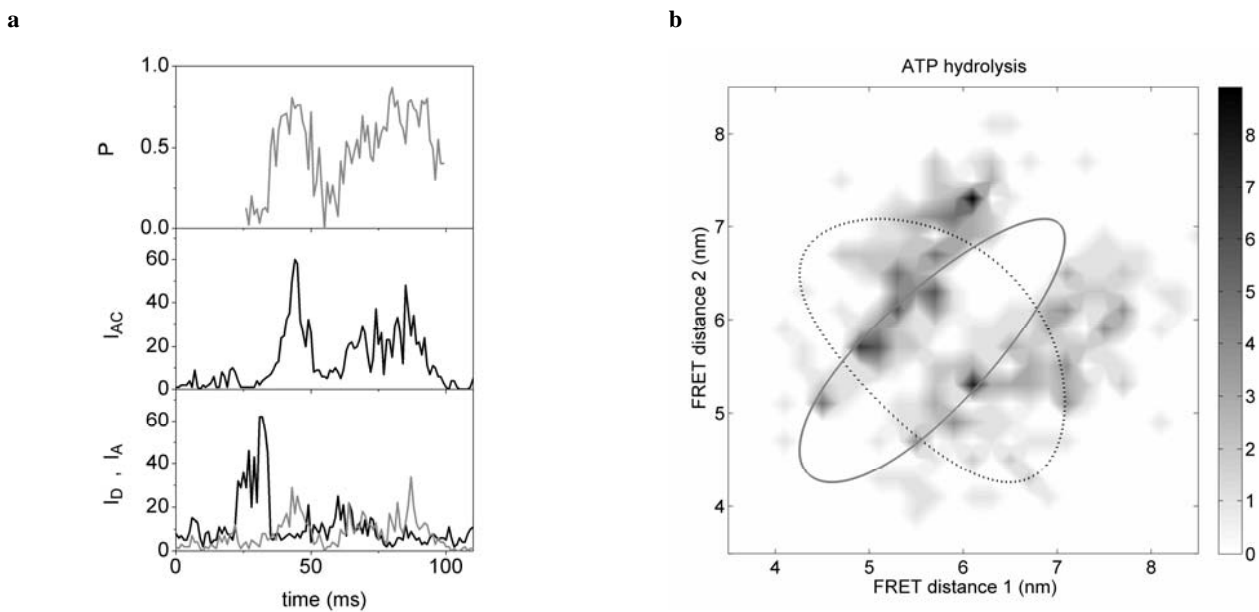

Fig. 6. **a**, confocal FRET time trajectories of a single freely diffusing $F_oF_1$-ATP synthase labeled with EGFP as FRET donor and Alexa568 as FRET acceptor in presence of 1 mM ATP. Lower panel, fluorescence intensities of EGFP ($I_D$, dark grey trace) and Alexa568 ($I_A$, grey trace) by 488 nm excitation. Middle panel, fluorescence intensity of Alexa568 using 561 nm excitation ($I_{AC}$, direct FRET acceptor probing). Upper panel, corresponding proximity factor trace P=$I_A/(I_A+I_D)$. **b**, FRET transitions density plot for all assigned FRET level transitions. The grey ellipse represents the expected FRET distance changes according to 36° stepping, the dotted curve represents 120° stepping.

### 3.2. FRET measurements of single $F_oF_1$-ATP synthases labeled at subunits ε and *c* during ATP hydrolysis

We labeled the ε subunit with Alexa532 as the FRET donor and the *c* subunit with Cy5 as the FRET acceptor to monitor changes between these positions. As shown for a single ATP synthase in Fig. 7a, the proximity factor P fluctuates in small steps within this photon burst indicating relative movements between these positions. About 40% of all photon bursts showed this proximity factor fluctuation both during ATP hydrolysis as well as ATP synthesis. In the absence of nucleotides, however, the percentage of bursts showing internal distance fluctuation was about 24%.

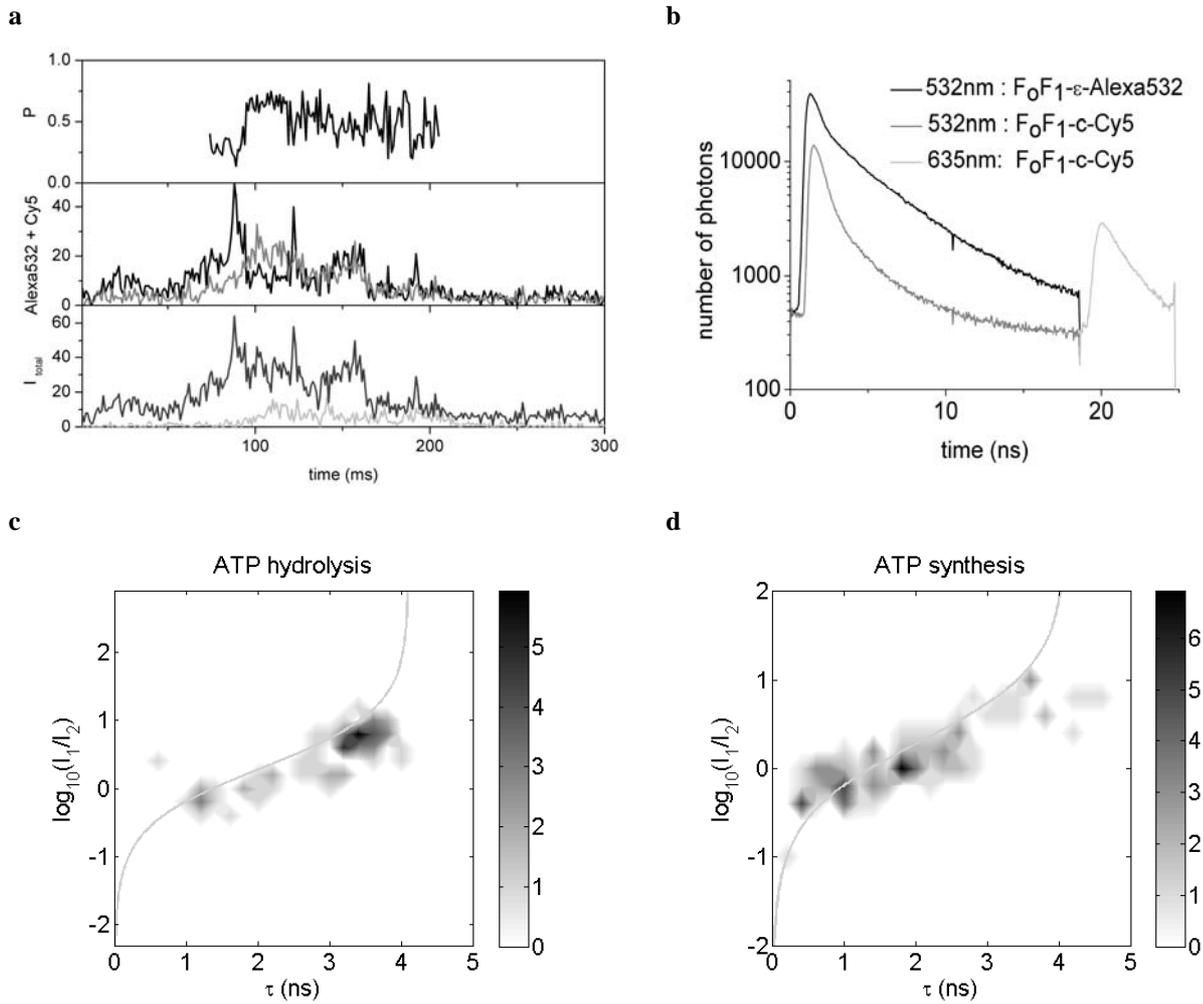

Fig. 7. **a**, FRET time trajectories of a single freely diffusing $F_oF_1$-ATP synthase labeled at ε with Alexa532 as FRET donor and with Cy5 as FRET acceptor at *c* in the presence of 1 mM ATP. Upper panel, proximity factor trace; middle panel, intensity trajectories for Alexa532 (black) and Cy5 (grey); lower panel, sum of both fluorescence intensities for 532 nm excitation (black) and fluorescence of Cy5 by direct excitation with 635 nm (grey). **b**, fluorescence lifetime histograms for pulsed excitation at 532 nm for FRET, or 635 nm for Cy5 only. **c**, FRET level plot of donor lifetime *versus* intensity ratio upon ATP hydrolysis. **d**, FRET level plot of donor lifetime *versus* intensity ratio upon ATP synthesis.

In order to exclude photophysical artifacts as the cause of the observed proximity factor changes we analyzed the FRET distance dependence using pulsed alternating laser excitation at 532 nm and 635 nm (Fig. 7b). Quantitative FRET distance information can be obtained from the ratio of the measured fluorescence intensities of FRET donor and acceptor

$$E_{FRET} = \frac{I_A}{\gamma I_D + I_A} \qquad (2)$$

(with $I_A$ and $I_D$, fluorescence intensities from the acceptor/donor dye, and γ, correction factor comprising quantum yields of the dyes and the detection efficiencies of both detection channels), or independently from the lifetime information of the FRET donor dye. An increasing FRET efficiency is reflected in a reduction of the donor lifetime:

$$E_{FRET} = 1 - \frac{\tau_{DA}}{\tau_{D_0}} \tag{3}$$

with $\tau_{DA}$, the fluorescence lifetime of the FRET donor in presence of the acceptor, $\tau_{D_0}$ is the fluorescence lifetime of the donor in the absence of a FRET acceptor or other quenchers in the local environment. As intensity ratio and lifetime are related by equations (2) and (3) FRET changes should fall on the lines drawn in Fig. 7 c and d, allowing discrimination from quenching of either FRET donor or acceptor, which would not fall on the lines. We plotted the intensity ratio *versus* lifetime for each assigned FRET level in photon bursts showing at least two distinct FRET levels, and we found a very good agreement with the predicted distance dependence according to FRET. The broad distribution of FRET donor lifetimes for these FRET levels is due to the fact that at least 5 distinct relative orientations, or distances for the fluorophores on ε and one of the 10 *c* subunits, have to be expected, with likely distances between 4 to 8 nm.

Thus, the smaller step sizes for *c* ring rotation in contrast to the well-defined 120° stepping of the γ and ε subunits must relate to internal elasticity of the rotor[46] and correspond to relative movements of the labeled sites on ε and *c*.

### 3.3. Triple-FRET measurements of single $F_oF_1$-ATP synthases labeled at subunits *a*, ε and *c*

The remaining question is how to verify that the internal elastic movements in the rotor subunits are observed while the whole rotor is turning. To make this determination, an auxiliary external static reference position is needed. We chose the non-rotating *a* subunit and fused the autofluorescent EGFP to the C terminus of *a*. In addition the previously used FRET labeling positions on subunits ε and *c* were labeled with Alexa532 (ε56) and Cy5 on *c* (see Fig. 3).

We monitored the rotation of the rotor with respect to the stator by FRET between EGFP on *a* and both FRET acceptors on ε and *c* using pulsed excitation at 488 nm. During ATP hydroylsis the rotor movements should occur in a stepwise manner because we expect a larger energy transfer to Alexa532 on the ε subunit which was shown to rotate in 120° steps. Simultaneously we detected the internal flexibility of the rotor by interleaved excitation of Alexa532 on the ε with 532 nm by switching an AOM. For this time interval, Alexa532 acted as a FRET donor with respect to the FRET acceptor Cy5 on *c*. A third interleaved laser pulse at 635 nm was applied to directly probe the existence of the Cy5 dye on *c* (see Fig. 5). However it turned out that the detectable fluorescence intensity from the EGFP as FRET donor 1 was too low using this pulse scheme. Therefore we continued with a modified 2-laser sequence for the triple FRET experiments starting with a series of 4 pulses at 488 nm, each delayed by 12 ns, followed by an 40 ns pulse at 532 nm within a 96 ns cycle.

One of the most prominent photon bursts of a single triply-labeled $F_oF_1$-ATP synthase in the presence of 1 mM ATP is shown in Fig. 8. The enzyme dynamics were recorded for more than 200 ms and the internal movements were classified into 5 phases:

- From the start up to about 48 ms the rotor was oriented in a low FRET efficiency orientation as seen in fluorescence intensities in the lowest panel d for EGFP (black trace) and the sum of the two FRET acceptor intensities (grey trace). At the same time the ε-c distance corresponds to a mean proximity factor P~0.4 in panel a.
- Between 48 ms and 58 ms the ε-c distance decreased slightly (see change in panel a denoted as P(3-4)) but an overall rotor movement cannot be deduced as measured by FRET from EGFP to the other dyes. The associated proximity factor P(1-2) remained constant in this second time interval, but moved afterwards.
- Between 58 ms and 89 ms the rotor was found in a medium FRET orientation with strong oscillations in the FRET donor intensity trace. For this time interval the internal rotor dynamics were not prominent and the ε-*c* distance seemed to be unchanged.
- Between 89 ms and 96 ms rotor had moved towards a high FRET orientation. At the same time the internal conformation and distance between the ε and *c* did not change.
- However, after 96 ms the internal ε-*c* distance changed to a higher proximity factor or shorter distance, and also the high FRET orientation of the rotor with respect to the *a* subunit seemed to change towards an even closer orientation.

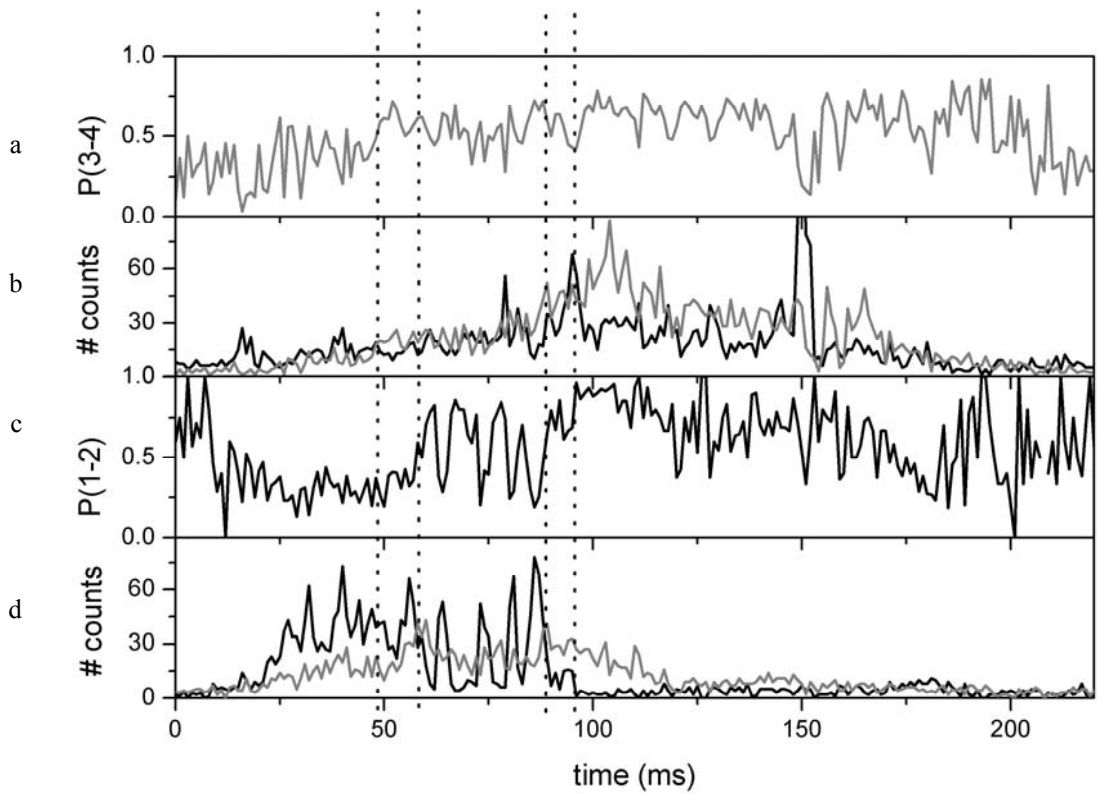

Fig. 8. Time trajectory of the fluorescence intensity changes within a single triply-labeled F$_o$F$_1$-ATP synthase in the presence of 1 mM ATP. **a**, proximity factor (P3-4) trace for FRET between Alexa532 on subunit ε and Cy5 on subunit *c* upon 532 nm excitation. **b**, fluorescence intensities of Alexa532 (black) and Cy5 (grey) upon 532 nm excitation. **c**, proximity factor (P1-2) tracce calculated from FRET donor EGFP and the sum of both FRET acceptors excited with 488 nm. **d**, fluorescence intensities for EGFP (black) and the sum of both FRET acceptors Alexa532 and Cy5 (grey) upon excitation with 488 nm.

## 4. DISCUSSION

F$_o$F$_1$-ATP synthase exhibits rotary subunit movements during ATP hydrolysis and ATP synthesis. In order to gain detailed information about the mechanics and mechanisms of the rotary motions, single enzymes have to be investigated. Here the single-molecule FRET approach to monitor subunit rotation by distance measurements within single membrane-embedded enzymes in real time had been refined. In order to study the elastic transient energy storage within the rotor subunits and to compare ATP hydrolysis with ATP synthesis coonditions, the doubly- or triply-labeled F$_o$F$_1$-ATP synthase were reconstituted in liposomes and investigated as freely diffusing proteoliposomes.

We have observed the conformational dynamics of single F$_o$F$_1$-ATP synthase using distance measurements between fluorophores on rotating and static subunits. The triple-FRET experiment shown here allows for simultaneous monitoring of two distinct types of conformational changes within a single enzyme, *i.e.* turnover-associated rotation and the internal dynamics of the rotor. The single-molecule approach can be extended to investigate conformational dynamics of the stator subunits b$_2$ with respect to rotation, to associate the effect of bound fluorescent inhibitors to rotation, to triangulate the positions of parts of the enzyme like the C terminus of the δ subunit, or to characterize nucleotide binding-dependent conformational changes in stator or rotor.

To overcome the limited observation time of the freely diffusing liposome with F$_o$F$_1$-ATP synthase a trapping scheme is required. In collaboration with W.E. Moerner (Stanford) and A.E. Cohen (Harvard) we started to refine their ABEL trap[47-52] approach (Anti-Brownian Electrokinetic Trap). In the so-called "software-based trap" version, a fluorescent nano-object or liposome is detected by a CCD camera while moving in a flat microfluidic chamber. Brownian motion is confined to two dimensions due the limited height of the chamber. An arbitrary reference x-y position is defined, and, depending on the actual distance of the liposome to this point, voltages are applied to four electrodes at the ends of the four channels leading towards the trapping region. Due to mass transport and an electric field component, the liposome is pushed back to the reference point. Using a fast CCD camera with single-molecule sensitivity, the ABEL trap restricts deviations from the reference position to less than 300 nm. Placing the confocal lasers for the single-molecule FRET experiments to this reference position will allow continuous monitoring of the conformational dynamics of a single $F_oF_1$-ATP synthase for a more extended period.

**Acknowledgements**

The authors want to thank Prof. Dr. P. Gräber and coworkers (University of Freiburg, Germany) for the gift of the Alexa532-labeled $F_1$ parts of ATP synthase which where used in the triple-FRET experiments. We thank Y. Bi and D. J. Cipriano (University of Western Ontario, London, Canada) for support in constructing the plasmids for EGFP-fused $F_oF_1$-ATP synthases. Financial support by the Deutsche Forschungsgemeinschaft (BO 1891/8-1 , BO 1891/10-1) and the Landesstiftung Baden-Württemberg (network of competence 'functional nanodevices') and the Canadian Institutes of Health (MT-10237) is gratefully acknowledged.